\documentclass{emulateapj}
\usepackage{apjfonts}
\usepackage{gensymb}
\usepackage{epsf}
\usepackage{graphicx}
\begin{document}
\slugcomment{}
\shortauthors{El-Batal et al.}
\shorttitle{GS 1354$-$645}

\title{NuSTAR Observations of the Black Hole GS 1354$-$645: Evidence
  of Rapid Black Hole Spin}

\author{A.~M.~El-Batal\altaffilmark{1},
J.~M.~Miller\altaffilmark{1},
M.~T.~Reynolds\altaffilmark{1},
S.~E.~Boggs\altaffilmark{2},
F.~E.~Chistensen\altaffilmark{3},
W.~W.~Craig\altaffilmark{2,4}
F.~Fuerst\altaffilmark{5},
C.~J.~Hailey\altaffilmark{6}
F.~A.~Harrison\altaffilmark{5},
D.~K.~Stern\altaffilmark{7},
J.~Tomsick\altaffilmark{2},
D.~J.~Walton\altaffilmark{7},
W.~W.~Zhang\altaffilmark{8}
}
 
\altaffiltext{1}{Department of Astronomy, University of Michigan, 1085
  S University Ave, Ann Arbor, MI 48109, USA, jonmm@umich.edu}

\altaffiltext{2}{Space Sciences Laboratory, 7 Gauss Way, University of
  California, Berkeley, CA 94720-7450, USA}

\altaffiltext{3}{Danish Technical University, Lungby, Denmark}

\altaffiltext{4}{Lawrence Livermore National Laboratory, Livermore, CA
  94550, USA}

\altaffiltext{5}{Cahill Center for Astronomy and Astrophysics,
  California Institute of Technology, 1200 East California Boulevard,
  Pasadena, CA 91125, USA}

\altaffiltext{6}{Columbia Astrophysics Laboratory and Department of
  Astronomy, Columbia University, 550 West 120th Street, New York, NY
  10027, USA}

\altaffiltext{7}{Jet Propulsion Laboratory, California Institute of
  Technology, 4800 Oak Grove Drive, Pasadena, CA 91109, USA}

\altaffiltext{8}{NASA Goddard Space Flight Center, Greenbelt, MD 20771, USA}

\keywords{accretion disks -- black hole physics -- X-rays: binaries}

\label{firstpage}

\begin{abstract}
We present the results of a {\it NuSTAR} study of the dynamically
confirmed stellar-mass black hole GS 1354$-$645.  The source was
observed during its 2015 ``hard'' state outburst; we concentrate on
spectra from two relatively bright phases.  In the higher-flux
observation, the broadband {\it NuSTAR} spectra reveal a clear, strong
disk reflection spectrum, blurred by a degree that requires a black
hole spin of $a = cJ/GM^2 \geq 0.98$ ($1\sigma$ statistical limits
only).  The fits also require a high inclination: $\theta \simeq
75(2)^{\circ}$.  Strong ``dips'' are sometimes observed in the X-ray
light curves of sources viewed at such an angle; these are absent,
perhaps indicating that dips correspond to flared disk structures that
only manifest at higher accretion rates.  In the lower-flux
observation, there is evidence of radial truncation of the thin
accretion disk.  We discuss these results in the context of spin in
stellar-mass black holes, and inner accretion flow geometries at
moderate accretion rates.
\end{abstract}

\section{Introduction}
Low-mass X-ray binaries are binary systems comprised of a compact object
accreting matter from a low-mass companion star.  The accretion disk
is luminous across the electromagnetic band pass, but it
peaks in the X-ray band.  The hot corona contributes in the hard X-ray
band.  Without a need for strong bolometric corrections, X-ray studies
of these sources can accurately constrain the geometry of the inner
accretion flow and its energetic properties (for a review of
stellar-mass black holes in such settings, see Remillard \& McClintock
2006).  Owing to their proximity, Galactic binaries are excellent
laboratories for probing the extreme gravitational effects of black
holes.

Astrophysical black holes can be fully described by their mass and
``spin'' (dimensionless angular momentum; $a = cJ/GM^{2}$, where $-1
\leq a \leq 1$).  The location of the innermost stable circular orbit
(ISCO) around the black hole depends on the spin parameter.  For a
Schwarzschild black hole (zero spin), the radius of the the ISCO is at
$R_{ISCO} = 6~GM/c^2$.  In an extreme Kerr black hole (maximally
spinning), $R_{ISCO} \simeq 1 GM/c^2$ (e.g., Bardeen, Press, \&
Teukolsky 1972).

Hard X-rays produced via Componization and/or sychrotron in the corona
are ``reflected'' from the accretion disk, and this effect can be used
to measure the spin of both stellar-mass and supermassive black holes
(for recent reviews, see, e.g., Miller 2007, Reynolds 2014, Miller \&
Miller 2015).  The reflection spectrum is ``blurred'' by the strong
Doppler shifts and gravitational red-shifts close to the black hole,
effectively tracing the radius of the ISCO and the spin of the black
hole.  Resolution is important in such studies, but sensitivity and a
broad spectral band pass are also very important.  {\it NuSTAR}
(Harrison et al.\ 2013) has unprecedented sensitivity in the 3-79 keV
band and does not suffer from distorting effects such as photon
pile-up; it is an ideal mission for studies of reflection and black
hole spin (see, e.g., Risaliti et al.\ 2013; Miller et al.\ 2013a,b; King
et al.\ 2014; Parker et al.\ 2014; Tomsick et al.\ 2014; Furst et al.\ 2015).

Other important measurements can be obtained through reflection
modeling.  Among these are the inclination of the innermost accretion
disk, though models can also measure the ionization of the inner disk,
and can also constrain elemental abundances.  In the best cases, the
size of the hard X-ray corona can even be constrained (e.g., Miller et
al.\ 2015).  Progress has not only relied upon improved X-ray
instrumentation, but also improved reflection models.  In particular,
\texttt{relxill} (e.g., Garcia et al.\ 2013, Dauser et al.\ 2014)
offers many advantages over prior models, including internal
relativistic blurring and angle-dependent scattering calculations.

The object of this work, GS 1354-645, is a binary system comprising a
dynamically confirmed black hole of mass $M_{BH} \geq ~7.6(7)~
M_\odot$ (Casares et al. 2009) and a low mass stellar companion.  It
was first detected using the All Sky Monitor (ASM) aboard the {\it
  Ginga} satellite during an outburst in 1987 (Kitamoto et
al.\ 1990). The last outburst of this source was detected using {\it
  RXTE} in 1997 (e.g. Revnivtsev et al.\ 2000, Brocksopp et al 2001;
also see Reynolds \& Miller 2013). The distance to the source is not
well-constrained, with estimates lying between 25 and 61 kpc (Casares
et al. 2009).

Monitoring with the {\it Swift}/BAT detected a new outburst of
GS~1354$-$645 in June 2015 (Miller, Reynolds, Kennea 2015).
Better-known, recurrent sources like GX~339$-$4 often show multiple
spectral states, but GS 1354$-$645 is interesting in that only the
``hard'' state was observed in its last outburst.  GS~1354$-$645
may therefore belong to a small sub-class of black hole transients
including the better-known GRO J0422$+$32 and XTE~J1118$+$480
(Brocksopp et al.\ 2001).  Disk reflection was clearly required in
{\it RXTE} spectra of the 1997 outburst of GS~1354$-$645 (Revnivtsev
et al.\ 2000), potentially indicating a means to study the black hole
and innermost accretion flow in a ``hard state'' transient.  We
therefore requested observations with {\it NuSTAR}.

\section{Data Reduction}
We obtained two {\it NuSTAR} observations of GS 1354-645 during its
2015 outburst.  The first observation was made on 2015 June 13
(hereafter, Obs.\ 1), starting at 06:56:07 (UT).  It achieved a net
exposure of 24~ks.  The second observation (hereafter, Obs.\ 2) was
made roughly one month later, on 2015 July 11, starting at 13:41:08
(UT).  Obs.\ 2 achieved a net exposure of 29~ks.

Data reduction was performed using the routines in HEASOFT version
6.16, and the associated {\it NuSTAR} calibration files (version
20150316).  The {\it nuproducts} routine was run to extract source
light curves, spectra, and responses, as well as background spectra.
FPMA and FPMB source events were extracted using a 148 arc second
(radius) circle, centered on the source.  Background events were
extracted from a region of equivalent size in a source-free
region.

\section{Analysis and Results}
Figure 1 shows the light curve of the entire outburst, based on public
monitoring observations made with the {\it Swift}/BAT.  In qualitative
terms, the outburst has a ``fast rise, exponential decay'' or FRED
profile.  The points at which Obs.\ 1 and Obs.\ 2 were made are
indicated.  Observation 2 was obtained at a flux $\sim$5 times higher
than Obs.\ 1; this is reflected in the signal-to-noise ratios (S/N) of
the respective resultant spectra.

The spectra were analyzed using XSPEC version 12.8.2 (Arnaud 1996).
Fits were made across the full {\it NuSTAR} band (3--79~keV).  Spectra
from the FPMA and FPMB were fit jointly, allowing a multiplicative
constant to act as a flux normalization factor.  The spectra from both
observations were grouped to require a minimum of 30 counts per bin,
to ensure the validity of the $\chi^{2}$ fit statistic (Cash 1979,
Gehrels 1986).  All errors reported in
this work reflect $1\sigma$ confidence limits.

The 3~keV lower energy limit of {\it NuSTAR} is not suited to
constraining the column density of the ISM along the line of sight to
a source, when its value is low or moderate.  A value of $N_{\rm H} =
7\times 10^{21}~{\rm cm}^{-2}$ is suggested by the HEASARC column
density tool, based on Dickey \& Lockman (1990).  This value was fixed
in all fits to the spectra of GS 1354$-$645, using the ``tbabs'' model
(Wilms et al.\ 2000).

We initically considered a canonical spectral model consisting of
separate additive components, $diskbb$ (Mitsuda et al.\ 1984) and a
$power-law$.  Neither observation requires the disk blackbody
component.  It is possible that the disk temperature is simply too low
to be detected in the {\it NuSTAR} band (see, e.g., Reis et al.\ 2010,
Reynolds \& Miller 2013).

However, the simple power-law component does not achieve a formally
acceptable fit to either spectrum.  The fit to the lower-sensitivity
spectrum from Obs.\ 1 measured a power-law index of $\Gamma =
1.369(4)$, but only achieved $\chi^{2}/\nu = 2006.50/1747 = 1.149$
(where $\nu$ is the number of degrees of freedom in the fit).  The
spectrum of Obs.\ 2 is steeper, with $\Gamma = 1.528(1)$ in this
simple fit, and a very fit statistic results owing to the
improved S/N: $\chi^{2}/\nu = 9904.9/2740 = 3.615$.

These simple fits are shown in Figure 2.  The data/model ratio from
each fit shows residuals that are consistent with disk reflection,
including Fe K emission and a Compton back-scattering excess peaking
in the 20--30~keV range.  The ratio from Obs.\ 2 is consistent with
blurred reflection from an inner disk that extends close to the black
hole, similar to the features observed in the low/hard state of GRS
1915$+$105 (see Miller et al.\ 2013).

We next considered fits with \texttt{relxill} (version 0.4c; Garcia et
al.\ 2013; Dauser et al.\ 2014).  This model includes the power-law
that illuminates the disk, and also incorporates a Kerr blurring
function to translate from the frame of the accretion disk to the
frame of the observer.  It is this function that measures the strong
Doppler shifts and gravitational red-shifts imprinted on the
reflection spectrum.

Spin is measured directly using \texttt{relxill}, within the bounds
$-1 \leq a \leq 1$.  The inclination of the {\it inner} disk -- not
necessarily the same as the binary system -- is also measured
directly.  For these parameters, values obtained in fits to Obs.\ 2
were assumed in fits to Obs.\ 1, owing to its lower sensitivity.
\texttt{Relxill} also allows the inner disk to deviate from the ISCO
for a given spin parameter, so we also allowed the parameter $R_{in}$
to vary (the model has a hard upper limit of $1000~R_{ISCO}$) in fits
to both observations.

The emissivity of the disk is described as a broken power-law in
radius (e.g., $J \propto r^{-q}$), giving three parameters: $q_{in}$,
$q_{out}$, and $R_{break}$.  Any corona with an energetic profile
defined by the underlying potential and linked to the disk is likely
to have a time-averaged emissivity that falls with radius, so we
require $q \geq 0$ in fits to both observations.  The limited
sensitivity of Obs.\ 1 does not permit more detail to be discerned, so
we simply fixed $q_{out} = q_{in}$, obviating the meaning of
$R_{break}$.  For Obs.\ 2, however, additional constraints can be
imposed based on theoretical and observational expectations.  Wilkins
\& Fabian (2012) have calculated the emissivity profiles expected for
idealized point source scenarios, including a ``lamp post'' geometry.
All calculations assume a disk that extends to the ISCO.  The
predicted profiles are complex and a power-law with a single break
radius is a relatively crude approximation.  Inner emissivity profiles
steeper than $q=3$ (Euclidean) are only expected in a Kerr geometry,
so we have enforced a $R_{break} < 6 GM/c^{2}$ in all fits.  The
sensitivity afforded by {\it NuSTAR} observations of stellar-mass
black holes also suggests very small inner disk radii, and steep emissivity
indices (e.g., Miller et al.\ 2013, 2015; Parker et al.\ 2014; Tomsick
et al.\ 2014).  

In addition, we measured the following continuum and reflection
parameters using \texttt{relxill}: the power-law index $\Gamma$, the
flux normalization of the model, the cut-off energy of the
power-law ($E_{\rm cut}$; a hard upper limit of 1000~keV is fixed
within the model), the reflection fraction ($f_{refl}$, the ratio of
reflected to incident flux), the iron abundance ($A_{\rm Fe}$; a hard
lower limit of $0.5\leq A_{\rm Fe}$ is fixed within the model), and
the ionization of the accretion disk ($\xi = L/nr^{2}$, where $L$ is
the ionizing luminosity, $r$ is the distance between the source and
reflector, and $n$ is the density of the reflecting medium).

The results of fits to the spectra obtained in Obs.\ 1 and Obs.\ 2
with \texttt{relxill} are detailed in Table 1, and shown in Figure 3.
Owing to the much higher S/N of Obs.\ 2, we fit this spectrum
first.  The black hole spin, inner disk inclination, and iron
abundance measured in Obs.\ 2 were then frozen in fits to Obs.\ 1.

Far better fits are achieved with \texttt{relxill} ($\chi^{2}/\nu =
2753.0/2730 = 1.008$ for Obs.\ 2, and $\chi^{2}/\nu = 1890.2/1742 =
1.085$ for Obs.\ 1).  Most importantly, a very high black hole spin
parameter is measured in Obs.\ 2: $a = 0.998_{-0.009}$.  With the
sensitivity of {\it NuSTAR}, statistical errors on spin can be small,
but systematic errors are likely much larger (see below).  The
inclination of the inner disk is also very tightly constrained: $i =
75(2)$~degrees.  Since spin and inclination both work to blur the
reflection spectrum, it is important to ensure that the values are not
degenerate.  Both parameters are tightly constrained, and there
appears to be no degeneracy.  The inner disk radius is found to be
consistent with the ISCO and tightly constrained: $R_{in} =
1.07^{+0.05}_{-0.02}~ R_{ISCO}$.  This again indicates a disk that
extends very close to the black hole.

In contrast to Obs.\ 2, fits with \texttt{relxill} indicate that the
accretion disk did not extend close to the black hole in Obs.\ 1.  The
inner disk radius hits the limit of the model, $R_{in} =
700_{-500}^{+200}~ R_{ISCO}$.  This is indicated by the much narrower
line in Obs.\ 1 (see Fig.\ 2 and Fig.\ 4).  The statistical certainty
of this result is low; an inner radius of 80~$R_{ISCO}$ is within the
90\% confidence range.  We note that the Fe K line is not ideally fit,
based on the data/model ratio in Figure 3; this may imply that the
disk is truncated at a larger radius than allowed within the relxill
framework.  The addition of a simple Gaussian with an energy fixed at
$E = 6.40$~keV improves the fit at the $3\sigma$ level of confidence.

Although the reflection fraction in Obs.\ 1 is lower than in Obs.\ 2
($f_{refl} = 0.50(7)$ versus $f_{refl} = 1.5(1)$), it is still fairly
high for a compact corona and a much larger inner disk radius.  A high
reflection fraction and a relatively {\it flat} emissivity index in
Obs.\ 1 may be nominally consistent with a large corona that partially
blankets the disk, but a standard emissivity profile ($q = 3$) is
allowed by the data at the 90\% confidence level.

\section{Discussion and Conclusions}
We have analyzed two {\it NuSTAR} spectra of the dynamically confirmed
black hole GS~1354$-$645.  Both observations were obtained in the
``low/hard'' state.  When fit with a
relativistic reflection model, the spectrum obtained close to the
outburst peak suggests a high spin parameter, and also implies
that the inner disk may be viewed at a high inclination.  The spectrum
obtained at a lower flux level suggests that the inner disk may have
been truncated, potentially consistent with radiatively
inefficient accretion flow models.  In this section, we discuss the
strengths and weaknesses of our results, potential sources of
systematic errors, and impacts on our understanding of GS 1354$-$645.

Fits to Obs.\ 2 with \texttt{relxill} indicate that the accretion disk
likely extends very close to the black hole.  This is now common in
the most luminous phases of the ``low/hard'' state, especially at the
sensitivity afforded by {\it NuSTAR} (see, e.g., Miller et al.\ 2015).
The data strongly require a very high spin parameter, $a =
0.998_{-0.009}$, consistent with the extreme upper bound of the model.
The error is merely the statistical error, and systematic errors are
likely to be much larger.

All measures of black hole spin obtained through the accretion disk
rely on the optically thick reflecting gas obeying the test particle
ISCO.  Simulations suggest that the disk is likely to be thin and to
obey the ISCO at Eddington fractions below 0.3 (Shafee et al.\ 2008,
Reynolds \& Fabian 2008); for plausible combinations of black hole mass
and distance, this condition was met in our observations.  It is quite
possible, however, that no astrophysical disk respects the ISCO at the
percent level.  

The best-fit model for Obs.\ 2 yields parameters similar to those
expected in a ``lamp-post'' geometry (a compact, central source of
hard X-rays located on the spin axis above the black hole).  The
\texttt{relxilllp} model (Garcia et al.\ 2013; Dauser et al.\ 2014)
explictly assumes this geometry and calculates the reflection fraction
self-consistently; it also gives a high spin parameter ($a > 0.85$).
However, our best-fit model (see Table 1) is superior at the
$6.6\sigma$ level of confidence, as determined by an F-test (for
\texttt{relxilllp}, $\chi^{2}/\nu = 2806/2733$, even if the reflection
fraction is not linked to the lamp-post value).  This might imply that
the corona is indeed compact but not quite an idealized lamp-post.
Recent work has noted some physical difficulties with idealized
lamp-post models (e.g., Niedzwiecki et al.\ 2016, Vincent et
al.\ 2016).

Stiele \& Kong (2016) have reported a nearly maximal retrograde spin
based on a short {\it XMM-Newton} observation of GS 1354$-$645.  A
combination of factors including calibration uncertainties in the
EPIC-pn camera and photon pile-up may have acted to falsely narrow the
reflection features in the {\it XMM-Newton} spectrum (Miller et
al.\ 2010).  It is also possible that the disk was mildly truncated
during the {\it XMM-Newton} observation.  Fits to {\it NuSTAR} Obs.\ 2
with $a \leq 0.93$ are rejected at the $5\sigma$ level of confidence,
via an F-test.

Our reflection modeling also indicates that the inner disk is viewed
at a relatively high inclination, $i = 75(2)$ degrees.  This is
within the eclipse limit derived by Casares et al.\ (2009).  It is
possible that the inner disk is not aligned with the inclination of
the binary system itself (Maccarone 2002; also see Tomsick et
al.\ 2014).

Systems that narrowly avoid eclipses are often observed to exhibit
``dips'' in their X-ray light curve.  These events may be due to
vertical structures in the outer accertion disk that block some of the
light from the central engine (see, e.g., Diaz-Trigo et al.\ 2006).
We did not detect any dips in the light curve of GS 1354$-$645,
possibly indicating that the inclination of the outer disk must really
be lower than the value derived for the inner disk via reflection.
However, dips may only be manifested at higher
Eddington fractions (see, e.g., Kuulkers et al.\ 2000) than the
luminosities inferred in our {\it NuSTAR} exposures.

Esin et al.\ (1997) predict that the inner accretion flow will become
advection-dominated and radiatively inefficient at Eddington fractions
below 0.01.  In the
0.01--0.08~$L_{Edd}$ range, the inner disk may still be truncated
but the inner flow can be more luminous. For GS 1354-645, assuming a
distance at the lower limit of $d=25$ kpc and mass at the lower limit
of $M = 7.6~M_{\odot}$, the luminosities based on the flux
values in Table 1 would be $0.1~L_{Edd}$ and $0.53L_{Edd}$
for Obs 1 and Obs 2 respectively. Smaller distances ($d<10$ kpc) would
more easily accommodate the lower end of the luminosity range at which
thin disks may truncate.  Alternatively, some combinations including
very high black hole masses ($M > 90~M_{\odot}$) can also meet
the prediction, but these prescriptions greatly exceed the range of
black hole masses inferred in X-ray binaries (e.g. Remillard \&
McClintock 2006).

We thank the anonymous referee for comments that improved this
manuscript.  This work was supported under NASA contract
No.\ NNG08FD60C, and made use of data from the {\it NuSTAR} mission, a
project led by the California Institute of Technology, managed by the
Jet Propulsion Laboratory, and funded by NASA.

\clearpage

\begin{figure}
\hspace{0.5in}
\includegraphics[scale=0.8]{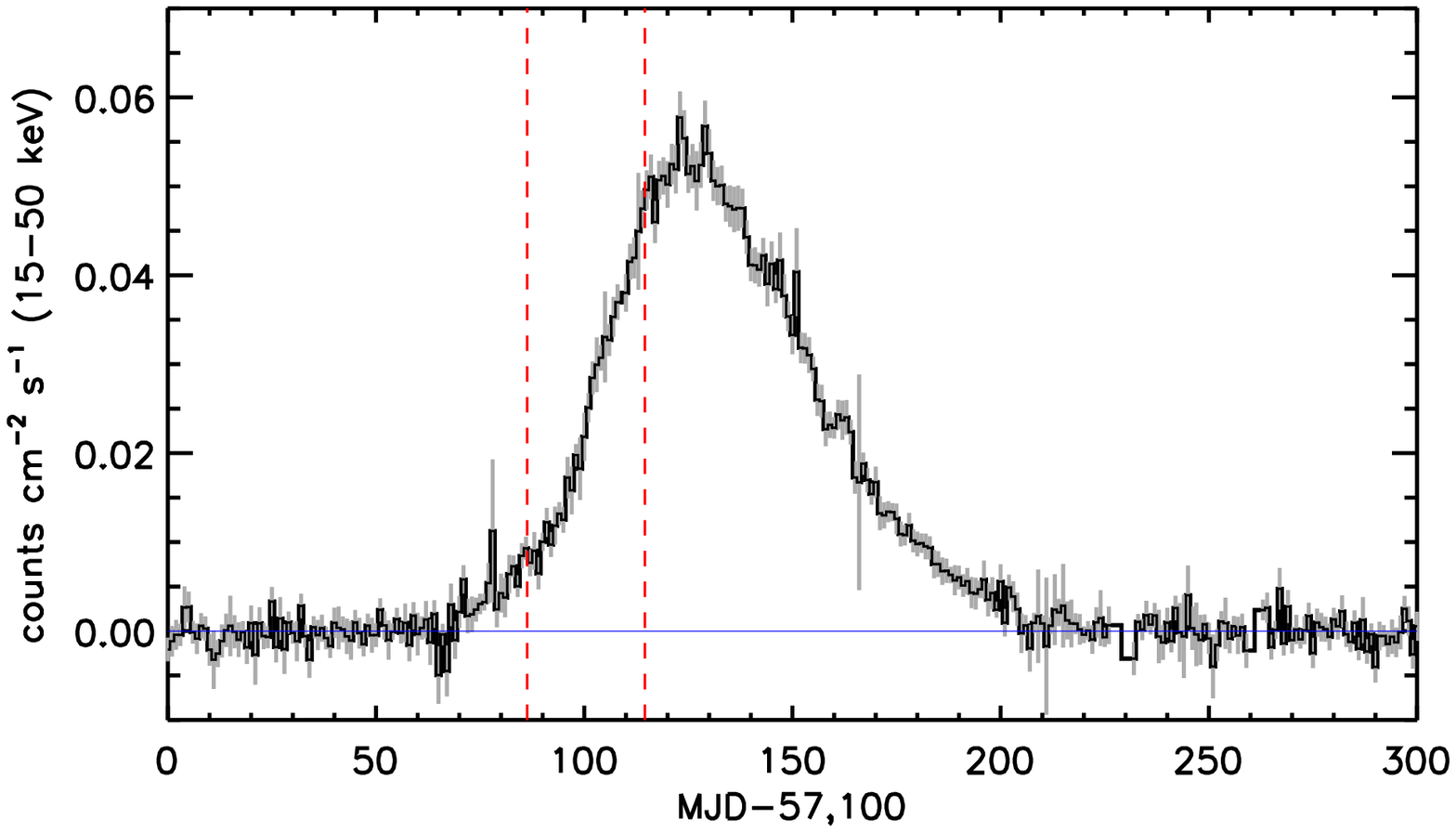}
\figcaption[t]{\footnotesize The {\it Swift}/BAT light curve of the
  2015 outburst of the black hole GS 1354$-$645.  The data points
  represent the average flux over a span of one day.  The red, dashed,
  vertical lines indicate the start time of the two {\it NuSTAR}
  observations considered in this work, referred to as Obs.\ 1 and
  Obs.\ 2.  The outburst is qualitatively consistent with a fast rise,
  exponential decay (FRED) profile.}
 \end{figure}

\clearpage

\begin{figure}
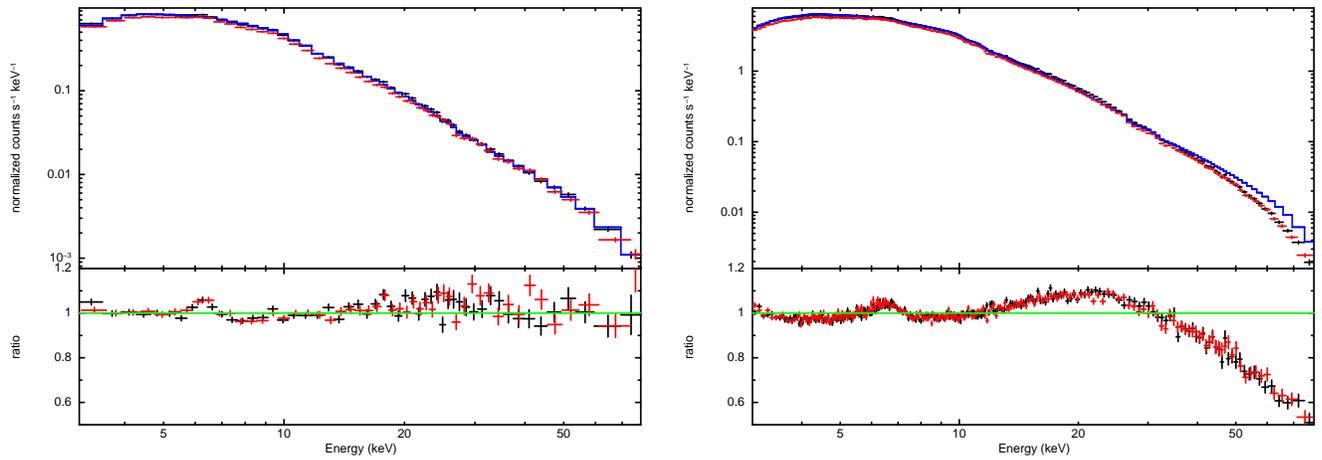

\includegraphics[scale=0.35,angle=-90]{f2a.ps}
\includegraphics[scale=0.35,angle=-90]{f2b.ps}
\figcaption[t]{\footnotesize {\it NuSTAR} spectra of GS~1354$-$645,
  fit with a fiducial power-law continuum.  The FPMA is shown in
  black, the FPMB is shown in red, and the model for the FPMA is shown
  in blue.  The data in each panel have been rebinned for visual
  clarity.  {\bf LEFT}: The spectrum of Obs.\ 1 is shown.  {\bf
    RIGHT}: The spectrum of Obs.\ 2 is shown.  This ratio reveals
  evidence of strong, broadened reflection consistent with
  illumination of the inner accretion disk.}
\end{figure}

\clearpage

\begin{table}[t]
\caption{Spectral Fitting Results}
\begin{footnotesize}
\begin{center}
\begin{tabular}{lll}
\tableline
\tableline
Parameter & Obs.\ 1 & Obs.\ 2 \\
\tableline
$q_{in}$  & $1(1)$ & $9(1)$  \\
$q_{out}$  & $=q_{in}$ & $0.0^{+0.4}$   \\
$R_{break}$ &   -- & $5(1)$   \\
$a~ ({\rm cJ/CM^{2}})$  & 0.998* & $0.998_{-0.009}$  \\
$i$~ (degrees) & 75 &  75(2)   \\
$R_{in}~ (R_{\rm ISCO})$ & $700_{-500}^{+100}$ & $1.07^{+0.05}_{-0.02}$  \\
$\Gamma$ & 1.46(1) & 1.635(7)  \\
${\rm log} (\xi)$ & 0.3(3) & $2.37(3)$  \\
$A_{\rm Fe}$ & 0.57* & 0.57(3)  \\
$E_{\rm cut}$ (keV) & $1000_{-200}$ & 150(5)  \\
$f_{\rm refl}$ & 0.50(7) & 1.5(1)  \\
${\rm Norm.}~(10^{-3})$ & 3.8(2) & 8.4(2)  \\
$\chi^{2}/\nu$ & 1890.2/1742 & 2753.0/2730  \\
\tableline
${\rm Flux}$ (3--79~keV, $10^{-9}~ {\rm erg}~ {\rm cm}^{-2}~ {\rm s}^{-1})$ &  1.01(1) &  5.6(1)    \\
${\rm Flux}$ (0.5--100~keV, $10^{-9}~ {\rm erg}~ {\rm cm}^{-2}~ {\rm s}^{-1})$ &  1.28(1) &  6.9(1)      \\
\tableline
\tableline
\end{tabular}
\vspace*{\baselineskip}~\\ \end{center} 
\tablecomments{The results of spectral fits to {\it NuSTAR}
  observations of the black hole X-ray binary GS 1354$-$645.  The
  ``relxill'' model (Dauser et al.\ 2014; Garcia et al.\ 2013) was
  used to describe the combination of the direct and reflected
  spectra.  The fits to Obs.\ 1 fix the parameters marked with an
  asterisk at the values measured in Obs.\ 2.  The inner disk
  inclination is described in terms of $i$.  The $R_{in}$ parameter is
  measured in units of the ISCO radius for the spin parameter $a$ and
  allows for the possibility that the disk is not truncated exactly at
  the ISCO.  Please see the text for details concerning the emissivity
  indices $q$ and break radius $R_{break}$.  The power-law index is
  noted by $\Gamma$.  The log of the ionization parameter, the
  abundance of Fe relative to solar, and the power-law cut-off energy
  are listed as ${\rm log}(\xi)$ , $A_{\rm Fe}$, and ${E}_{\rm cut}$,
  respectively.  The reflection fraction and flux normalization of
  ``relxill'' are listed as $f_{\rm refl}$ and ${\rm Norm.}$,
  respectively.  Last, the unabsorbed flux in the spectral fitting
  band, as well as a broader band, are listed based on these fits.}
\vspace{-1.0\baselineskip}
\end{footnotesize}
\end{table}
\medskip


\clearpage

\begin{figure}
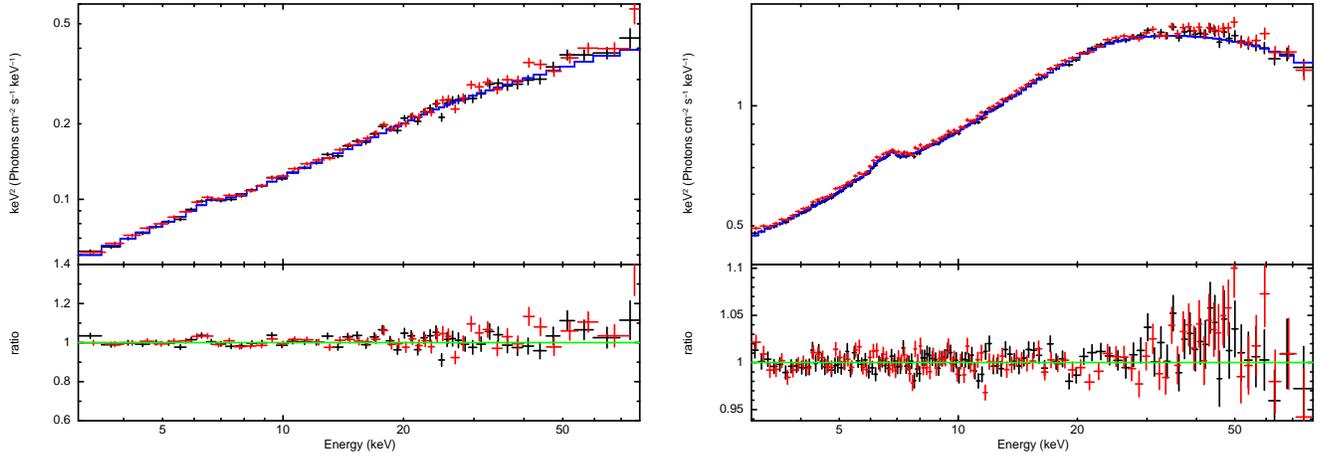

\includegraphics[scale=0.35,angle=-90]{f3a.ps}
\includegraphics[scale=0.35,angle=-90]{f3b.ps}
\figcaption[t]{\footnotesize {\it NuSTAR} spectra of GS~1354$-$645,
  fit with the ``relxill'' relativistically blurred disk reflection
  model.  The FPMA is shown in black, the FPMB is shown in red, and
  the model for the FPMA is shown in blue.  The data in each panel
  have been rebinned for visual clarity.  {\bf LEFT}: The spectrum of
  Obs.\ 1 is shown.  The data are consistent with reflection from a
  radially truncated accretion disk.  {\bf RIGHT}: The spectrum of the
  brighter Obs.\ 2 is shown.  The best fit requires a very high black
  hole spin parameter.}
\end{figure}

\begin{figure}
\hspace{-0.2in}
\includegraphics[scale=0.95]{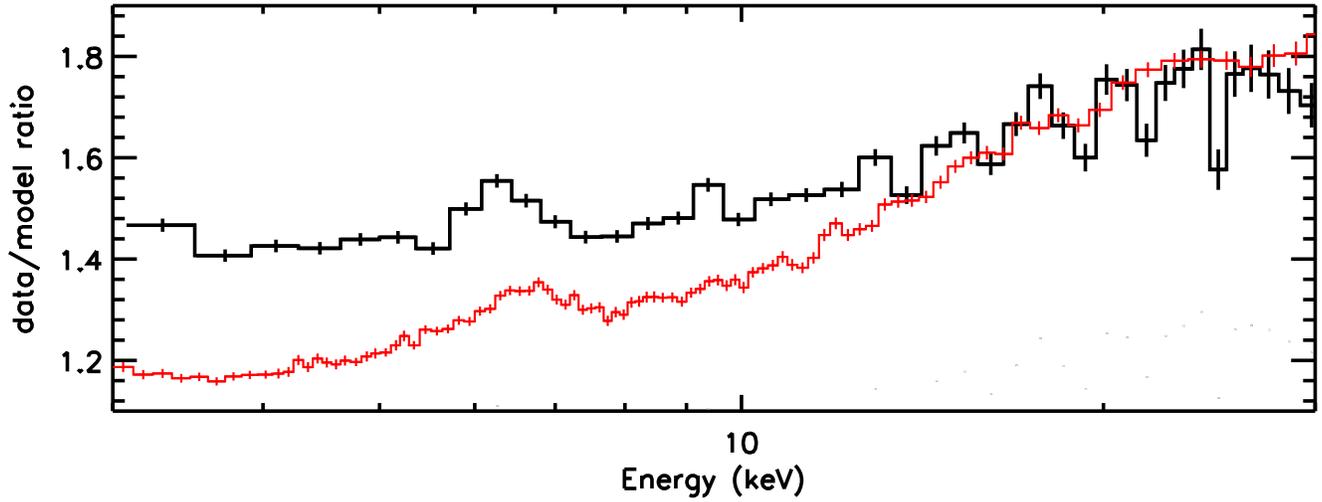}
\figcaption[t]{\footnotesize {\it NuSTAR} spectra from Obs.\ 1 (black)
  and Obs.\ 2 (red), after removing blurred disk reflection from the
  models detailed in Table 1.  The unmodeled Fe K line in Obs.\ 2
  extends down to 4--5~keV, consistent with a disk close to the ISCO
  around a spinning black hole.  In contrast, the unmodeled Fe K line
  in Obs.\ 1 is relatively narrow.  Only the FPMA spectra are shown,
  the ratio from Obs.\ 1 is multiplied by a factor of 1.4, and the
  data were binned for visual clarity.}
\end{figure}

\end{document}